\documentclass[11pt]{article}
\usepackage{tr-classic}
\usepackage{pgfplots}
\pgfplotsset{compat=1.18}

\title{Multimodal Reranking Is a System One Decision}
\reportshorttitle{Multimodal Reranking Is a System One Decision}
\reportlabel{}
\reportversion{}
\reportauthors{Zhuchenyang Liu, Yao Zhang, Yu Xiao}
\reportaffils{Aalto University}
\reportlinks{\textbf{Code:} \href{https://github.com/Ryenhails/rerank-as-decision}{github.com/Ryenhails/rerank-as-decision} \quad \textbf{Contact:} \href{mailto:zhuchenyang.liu@aalto.fi}{zhuchenyang.liu@aalto.fi}}
\reportabstract{%
Multimodal rerankers are usually dedicated models: a vision-language model (VLM) is trained on retrieval data to score
the candidates of a query. Scoring options is not specific to reranking. System One decision models such as Jev map a
state and a typed question to one probability per option without generating text, and general models can make such
decisions with little or no training. Reranking is one such decision, with the query and its candidates as the state
and the candidates as the options. We therefore argue that a general VLM can serve as a multimodal reranker through
System One decisions, and that decision tuning can improve its reranking. We ask a VLM one choice question over the top
16 candidates of a first-stage retriever and rank the candidates by the probabilities of their labels, read from one
forward pass. On MMEB-V2 image retrieval, this choice scores 5.4 to 7.3 hit@1 above a yes/no question per candidate,
and the yes/no question falls below the first stage. With the choice, zero-shot Qwen3.8-27B beats the supervised
Qwen3-VL-Reranker-2B on all three benchmarks and comes within 1.6 points of Reranker-8B on each. It matches
Reranker-8B on MMEB-V2 visual documents (84.79 vs.\ 84.78 nDCG@5), at 1.1 to 1.5 times its latency. JEV-27B-VL, the
same backbone tuned on text decisions only, improves page reranking by 0.6 to 1.1 points, with gains on most tasks
(sign test, $p \le 0.017$), and leaves natural images unchanged. The zero-shot model is larger than the supervised
rerankers and of a newer generation, so these comparisons are confounded; we discuss this and other alternative views,
and the experiments that would test the position.}

\begin{document}
\maketitle

\section{Introduction}
\label{sec:intro}

\begin{figure}[t]
\centering
\includegraphics[width=\linewidth]{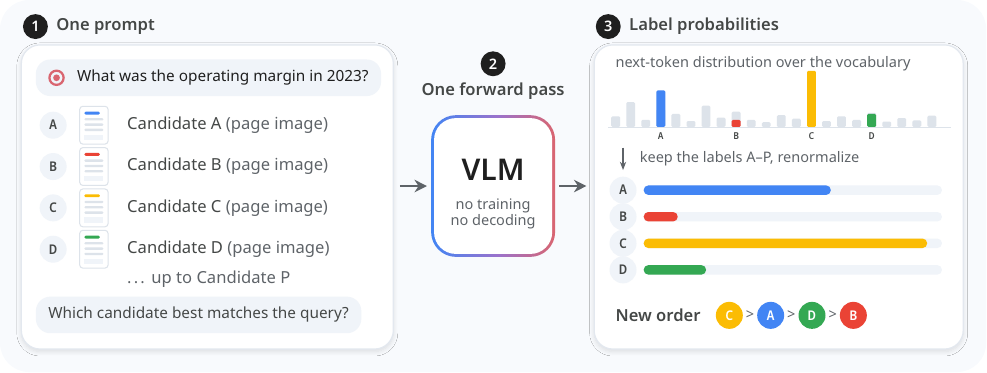}
\caption{The choice readout. (1) The query and the top-$K$ candidates of a first-stage retriever go into one prompt
under the labels A--P. (2) One forward pass of a vision-language model, with no decoding. (3) The next-token
distribution, restricted to the label tokens and renormalized, gives one probability per candidate, and the candidates
are reordered by it. Bar lengths are illustrative.}
\label{fig:method}
\end{figure}

Multimodal retrieval usually ranks in two stages. A first-stage retriever encodes the query and each item of the
corpus, such as an image or a page, independently into dense vectors and returns the most similar items as candidates.
A reranker then reorders the top candidates. Multimodal rerankers are usually dedicated models: a vision-language model
(VLM) is trained on retrieval data to read the query and one candidate together and output a relevance score, for
example the probability of ``yes''~\cite{li2026qwen3vlemb,jina2025m0}, or to rank several candidates at
once~\cite{liu2025lamra,sun2026ziprerank,cai2026ranknexus,lighton2026rerank}. In this view, better reranking means
training a better reranker, and a retrieval system serves the reranker as a separate model.

Reading a score from the model's output, rather than generating text, is not specific to reranking. System One decision
models such as Jev~\cite{almeida2026systemone,deusser2026jevbench} map a state and a typed question to one probability
per option, and are used for decisions of many kinds. General models can make these decisions: open language models
serve as Jev-style decision models with little or no training, by reading the probabilities of option labels from the
next-token distribution~\cite{li2026llmasjev,geuter2026innerjev,zhang2026anyjev,deusser2026jevbench}, and VLMs make
the same kind of decision when the state contains an image~\cite{zhou2026pixeljev,autotrust2026jevvl}.
Decision tuning, as in JEV-27B-VL~\cite{autotrust2026jevvl}, trains a general model to make such decisions across
tasks.

Reranking is such a decision: the state holds the query and its candidates, the question asks which candidate best
matches the query, and the options are the candidates. We therefore take the position that \textbf{a general VLM can
serve as a multimodal reranker through System One decisions, and that decision tuning can improve its reranking}.

The position matters for two reasons. First, for deployment: retrieval-augmented systems often already serve a large
VLM to generate answers. If that model can rerank through one decision over the candidates, the system needs no
separate reranker to train, serve and keep up to date. Second, for research: if reranking is one decision among many,
it can improve with decision tuning across tasks, without reranking-specific training.

The readout itself is not new: setwise prompting~\cite{zhuang2024setwise}, FIRST~\cite{reddy2024first} and
ZipRerank~\cite{sun2026ziprerank} all rank candidates by the probabilities of their labels. What is open is whether a
general VLM can rerank this way without training for it. ZipRerank reports that zero-shot VLMs of 8B and 11B
parameters, prompted as in RankGPT, do not consistently improve over the first stage~\cite{sun2026ziprerank}, and
zero-shot pointwise MLLM rerankers have helped only some query types~\cite{lin2025mmembed} or not at
all~\cite{erzurumlu2026geoloc}. In our experiments the answer depends on the question type and on scale. One choice
over all candidates improves the first stage on image retrieval, where a yes/no question per candidate falls below it.
Among Gemma~4 models, those of 12B parameters and more improve page retrieval zero-shot, and the two smaller ones lower
it (\cref{sec:an-scale}).

We test the position on MMEB-V2 image and visual-document retrieval~\cite{meng2025vlm2vecv2} and on
ViDoRe~V3~\cite{loison2026vidorev3}, reranking the top 16 candidates of a first-stage retriever. We first confirm on
MMEB-V2 image retrieval that the ``choice'' format (\cref{fig:method}) outperforms the ``noul'' (yes/no) format asked
zero-shot (\cref{sec:prelim}), and use it in all following experiments. Two findings support the position. First,
through one zero-shot decision, Qwen3.8-27B~\cite{qwen2026qwen38} beats a dedicated 2B reranker on all three
benchmarks, comes within 1.6 points of a dedicated 8B reranker on each, and matches it on MMEB-V2 visual documents.
Second, the same VLM tuned on text decisions alone reranks pages slightly better on most tasks, and natural images no
better.

These findings are encouraging but confounded. The VLM we test is larger than the dedicated rerankers and from a newer
model generation, so whether a dedicated reranker built on the same backbone would do better remains open.
\Cref{sec:alt,sec:test} discuss this and other alternative views, and the experiments that would test the position. We
release the code.

\section{Related Work}
\label{sec:related}

\paragraph{System One models}
TypeSafe introduced System One models with Jev, describing Jev as ``unstructured state in, typed probabilistic decisions
out'' and training it with reinforcement learning for calibrated decisions~\cite{almeida2026systemone}; its weights and
training method are not public~\cite{geuter2026innerjev}. Independent evaluations measured Jev on public text benchmarks~\cite{deusser2026jevbench}
and in applications~\cite{rafe2026benchmarking,ibrahim2026css}. Open reproductions read the same kind of
decision from frozen language models: LLM-as-Jev scores numbered options with a prefilled answer prefix~\cite{li2026llmasjev},
Releasing the Inner Jev reads letter logits at the last position and adds a light LoRA recipe~\cite{geuter2026innerjev},
and AnyJev divides out a label prior and averages over cyclic option rotations~\cite{zhang2026anyjev}. Deußer
et al.\ scored our backbone, Qwen3.8-27B, with this readout on text tasks~\cite{deusser2026jevbench}.

In these works the options are mostly text. PixelJev~\cite{zhou2026pixeljev}
answers questions about an image with text options. Where the options are images, the model picks among a few: the developers of JEV-27B-VL
use it to compare two generated images and to rank up to eight short videos for a recommendation
feed~\cite{autotrust2026jevvl,jev27bdemo2026,autotrust2026rankingteam}. Jev-style models have also been
applied to text retrieval: re-ranking TREC-COVID~\cite{autotrust2026jevvl}, choosing among
passages~\cite{chen2026auditretrieval}, and labelling data for a passage reranker~\cite{sadoun2026jevsearch}. We rerank
the images and pages returned by a retrieval system, on retrieval benchmarks, and compare with supervised multimodal
rerankers at the same depth.

\paragraph{LLM rerankers that read logits}
Zero-shot listwise reranking with LLMs started by generating permutations of passage
identifiers~\cite{sun2023rankgpt,ma2023lrl,pradeep2023rankvicuna,pradeep2023rankzephyr}. Setwise prompting showed that a
zero-shot model can rank a small window of passages by the likelihood of their labels~\cite{zhuang2024setwise}, and
FIRST trained a listwise reranker to rank by the logits of the first identifier token~\cite{reddy2024first,chen2024firstrepro}.
In-context reranking reads attention weights~\cite{chen2025icr,chen2026selectiveicr}. Pointwise rerankers read the
probability of a relevance label~\cite{nogueira2020monot5,zhuang2024beyondyesno}. In recommendation, listwise rerankers
that read identifier log-probabilities show position bias~\cite{bito2026positionbias}.
Our readout is that of setwise prompting and FIRST; we apply it zero-shot to images and pages.

\paragraph{Multimodal rerankers}
Supervised multimodal rerankers score each pair with a yes/no or relevance head~\cite{li2026qwen3vlemb,jina2025m0,lighton2024monoqwen}
or rank several candidates jointly~\cite{liu2025lamra,cai2026ranknexus,xu2025mmr5,lighton2026rerank}. Zero-shot MLLM rerankers have been
used pointwise, where they help multimodal queries and hurt single-modal ones~\cite{lin2025mmembed} or do not help at
all~\cite{erzurumlu2026geoloc}, and listwise by generating a permutation~\cite{eltahir2025vic,wu2025square}. Closest to us, ZipRerank trains a VLM to
rank pages by single-token letter logits in one forward pass~\cite{sun2026ziprerank}; its smaller zero-shot
VLM baselines (Llama-3.2-11B-Vision, Qwen3-VL-8B-Instruct) ``do not consistently improve over the first-stage
retriever''. We use the same readout without training for reranking, on natural images as well as pages: zero-shot on
larger VLMs than these baselines, and with a decision head tuned on text decisions only (\cref{sec:twoways}).

\section{Reranking as a System One Decision}
\label{sec:method}

\paragraph{Terms}
For decisions we use the terms of Jev~\cite{deusser2026jevbench,li2026llmasjev,autotrust2026jevvl}. A request has a ``state'', the context to judge,
and a typed ``question''. A ``choice'' question has a set of ``options'' and returns one probability per
option; a ``noul'' question is a yes/no question, answered with probabilities for ``false'' and ``true''. For retrieval we use
``query'' and ``candidates'', the items a first-stage retriever returns for a query.
\Cref{tab:mapping} shows how a reranking decision fills a Jev request under the two readouts we compare.

\begin{table}[h]
\centering\small
\caption{A reranking decision as a Jev request. Both readouts use the same instruction, query and candidates.}
\label{tab:mapping}
\begin{tabularx}{\linewidth}{lXX}
\toprule
Jev & Choice readout (listwise) & Yes/no readout (pointwise) \\
\midrule
question type & ``choice'' & ``noul'' \\
state & instruction, query, and the $K$ candidates under the labels A, B, \dots & instruction, query, and one candidate \\
question & ``Which candidate best matches the query, following the instruction?'' & ``Does this candidate match the query, following the instruction?'' \\
options & the $K$ candidate labels & none (a yes/no question) \\
output & one probability per candidate, summing to one & $P(\mathrm{yes})$ for the candidate \\
requests per query & 1 & $K$ \\
\bottomrule
\end{tabularx}
\end{table}

\subsection{Setting}
A first-stage retriever returns a ranked list for query $q$. We rerank its top $K$ candidates $c_1,\dots,c_K$
($K \le 16$), each an image, a page, a text, or an image with text. Candidates below rank $K$ keep the first-stage
order. Each benchmark supplies a task instruction (\cref{tab:instructions}).

\subsection{Choice readout}
\label{sec:choice}
The candidates enter the state in first-stage order, each introduced by its label, from ``Candidate A:'' to
``Candidate P:''. For an untuned VLM, the state and the question form one chat prompt with thinking disabled
(prompt~A in \cref{app:prompts}). One forward pass gives the next-token logits $z$; restricting them to the label
tokens $\ell_1,\dots,\ell_K$ gives
\begin{equation}
  p(c_j \mid q, c_{1:K}) = \frac{\exp z_{\ell_j}}{\sum_{k=1}^{K}\exp z_{\ell_k}},
  \label{eq:choice}
\end{equation}
and the candidates are ranked by $p$.

\subsection{Yes/no readout}
\label{sec:yesno}
The candidate is scored by the probability of ``yes'' against ``no'' at the first answer token, summing both casings
of each (\cref{app:impl}). This is the readout of supervised pointwise rerankers such as
Qwen3-VL-Reranker~\cite{li2026qwen3vlemb}, applied zero-shot.

\subsection{Two ways to obtain the probabilities}
\label{sec:twoways}
We obtain the probabilities in two ways on the same backbone, Qwen3.8-27B~\cite{qwen2026qwen38}, with the same state
and question (\cref{app:prompts}).

\paragraph{Label logits of an untuned VLM}
Qwen3.8-27B as released, with thinking disabled, read as in \cref{sec:choice,sec:yesno}; no weight is trained. We call
this ``no decision tuning''.

\paragraph{A trained decision head}
JEV-27B-VL~\cite{autotrust2026jevvl} is a Jev-style decision model built on the same backbone: its developers add a
LoRA adapter and a decision head to Qwen3.8-27B and train them on text decisions, leaving the vision encoder unchanged
(\cref{app:impl}). We call this ``text decision tuning''. No image enters the training, so every image decision
JEV-27B-VL makes, ours included, is zero-shot for the decision head.

\medskip\noindent
The two models share every weight except the adapter and the head. Comparing them isolates text decision tuning, and
comparing the two readouts on either model isolates the question type.

\section{Experimental Setup}
\label{sec:setup}

\paragraph{Benchmarks}
MMEB-V2~\cite{meng2025vlm2vecv2} image retrieval (12 tasks, hit@1) and visual-document retrieval (24 tasks,
nDCG@5), with the task list, data loaders and metrics of the Qwen3-VL-Embedding evaluation code; ViDoRe~V3~\cite{loison2026vidorev3}
(8 datasets, nDCG@10). For ViDoRe~V3 we pool the queries of all six languages within each dataset and average over
datasets. MMEB-V2 also has video retrieval tasks; we leave them out because JEV-27B-VL accepts no video input.

\paragraph{First stage}
Qwen3-VL-Embedding-2B~\cite{li2026qwen3vlemb}. All rerankers receive the same top-16 candidates.

\paragraph{Models}
Qwen3.8-27B and JEV-27B-VL are the controlled pair of
\cref{sec:twoways}; both run the choice readout on every benchmark and the yes/no readout on MMEB-V2 image retrieval.
The dedicated references are Qwen3-VL-Reranker-2B and -8B~\cite{li2026qwen3vlemb}, trained for reranking with a
pointwise yes/no score.

\paragraph{Image budget}
All models receive each image at most 1800 visual tokens ($1800\times32\times32$ pixels), the budget of the Qwen
rerankers.

\paragraph{Inference}
All runs use one NVIDIA H200; \cref{app:impl} gives the serving details.

\section{Preliminary: Choice vs.\ Yes/No Format}
\label{sec:prelim}

\begin{table}[h]
\centering\small
\caption{Choice vs.\ yes/no readout on MMEB-V2 image retrieval (12 tasks, hit@1, top 16 of Qwen3-VL-Embedding-2B).
The choice readout asks one question over the 16 candidates, the yes/no readout one question per candidate.
JEV-27B-VL answers both through its decision head. We ran the yes/no readout on image retrieval only.}
\label{tab:2x2}
\begin{tabular}{lccc}
\toprule
 & choice (1 pass) & yes/no (16 passes) & choice $-$ yes/no \\
\midrule
none (first stage) & \multicolumn{2}{c}{74.85} & \\
\midrule
Qwen3.8-27B (no decision tuning) & 76.76 & 69.50 & +7.26 \\
JEV-27B-VL (text decision tuning) & 76.80 & 71.42 & +5.38 \\
\midrule
JEV-27B-VL $-$ Qwen3.8-27B & +0.04 & +1.92 & \\
\bottomrule
\end{tabular}

\end{table}

\Cref{tab:2x2} crosses the two readouts with the two models on MMEB-V2 image retrieval. Asked one choice over the 16
candidates, Qwen3.8-27B and JEV-27B-VL reach 76.76 and 76.80 hit@1, above the first stage (74.85). Asked one yes/no
question per candidate, they reach 69.50 and 71.42, below it. The choice is higher on all 12 tasks for both models
(\cref{tab:pertask-image}), and the yes/no readout falls below the first stage on 11 tasks for Qwen3.8-27B and on 9
for JEV-27B-VL. \Cref{sec:why-choice} discusses why.

\begin{keyfindings}[Preliminary]
Of the two System One question types, one choice over all candidates improves the first-stage ranking on image
retrieval, and a yes/no question for each candidate, the form of supervised pointwise rerankers, makes it worse.
\end{keyfindings}

\section{Findings}
\label{sec:evidence}

The two findings test the two parts of the position: a general VLM reranks through System One decisions
(Finding~1), and decision tuning can improve that reranking (Finding~2). Both use the choice format of \cref{sec:prelim}.

\subsection{Finding 1: As a System One decision, a general VLM competes with dedicated rerankers}
\label{sec:ev-supervised}

\begin{table}[h]
\centering\small
\caption{Reranking the top 16 of Qwen3-VL-Embedding-2B. MMEB-V2: mean over tasks; ViDoRe~V3: mean over datasets, all
languages. Both 27B rows use the choice readout; the last row is their difference. Bold: best reranker in the column,
underlined: second best.}
\label{tab:main}
\begin{tabular}{lccc}
\toprule
Reranker & MMEB image & MMEB visdoc & ViDoRe V3 \\
 & hit@1 & nDCG@5 & nDCG@10 \\
\midrule
none (first stage) & 74.85 & 79.20 & 53.01 \\
Qwen3-VL-Reranker-2B (supervised) & 74.49 & 82.72 & 59.18 \\
Qwen3-VL-Reranker-8B (supervised) & \textbf{78.36} & 84.78 & \underline{61.98} \\
\midrule
Qwen3.8-27B, choice (no decision tuning) & 76.76 & \underline{84.79} & 60.94 \\
JEV-27B-VL, choice (text decision tuning) & \underline{76.80} & \textbf{85.42} & \textbf{62.03} \\
\midrule
JEV-27B-VL $-$ Qwen3.8-27B & +0.04 & +0.63 & +1.09 \\
\bottomrule
\end{tabular}

\end{table}

Zero-shot, Qwen3.8-27B is above Qwen3-VL-Reranker-2B on all three benchmarks (\cref{tab:main}), by 2.3
hit@1, 2.1 nDCG@5 and 1.8 nDCG@10. It matches Reranker-8B on MMEB-V2 visual documents (84.79 vs.\ 84.78), where it is
higher on 9 of the 24 tasks (\cref{tab:pertask-visdoc}). Reranker-8B leads by 1.6 hit@1 on natural images and by 1.0
nDCG@10 on ViDoRe~V3, where Qwen3.8-27B is higher on energy only (\cref{tab:pertask-vidore}). JEV-27B-VL is the best
reranker on both visual-document benchmarks (85.42 nDCG@5 and 62.03 nDCG@10).

\begin{keyfindings}[Finding 1]
Reranking through one System One decision, with no training for reranking, a general VLM beats the smaller dedicated
reranker on every benchmark and matches the larger one on MMEB-V2 visual documents.
\end{keyfindings}

\subsection{Finding 2: Text decision tuning improves reranking of pages}
\label{sec:ev-transfer}

The last row of \cref{tab:main} compares JEV-27B-VL with Qwen3.8-27B under the choice readout. On natural images the
two tie, with JEV-27B-VL higher on 4 of the 12 tasks. On pages JEV-27B-VL is higher by 0.63 nDCG@5 on MMEB-V2 visual
documents (17 of 24 tasks, two ties) and by 1.09 nDCG@10 on ViDoRe~V3 (all 8 datasets); a two-sided sign test over tasks gives $p = 0.017$ and
$p = 0.008$. With the yes/no readout, tuning adds 1.92 hit@1 on images (\cref{tab:2x2}). A plausible reason is that
pages are mostly text rendered as an image, the modality the tuning data covers.

\begin{keyfindings}[Finding 2]
Decision tuning on text alone slightly improves the VLM's reranking of document pages and leaves natural images
unchanged.
\end{keyfindings}

\section{Additional Experiments}
\label{sec:additional}

\subsection{Gemma~4 models}
\label{sec:an-scale}

\begin{figure}[t]
\centering
\begin{tikzpicture}
\begin{axis}[width=0.6\linewidth, height=6.2cm, xmin=0, xmax=34, ymin=-14, ymax=11,
  xtick={5.12,8.00,11.96,25.81,31.27}, xticklabels={{E2B\\5.1B},{E4B\\8.0B},{12B\\12.0B},{26B-A4B\\25.8B},{31B\\31.3B}},
  xticklabel style={align=center, font=\sffamily\scriptsize}, yticklabel style={font=\sffamily\scriptsize},
  xlabel={model size (total parameters)}, ylabel={gain over the first stage (points)},
  label style={font=\sffamily\footnotesize}, ytick={-12,-8,-4,0,4,8}, ymajorgrids, grid style={jevline},
  axis line style={jevsubtle}, tick style={jevsubtle},
  legend style={at={(1.03,0.5)}, anchor=west, font=\sffamily\scriptsize, draw=jevline2, rounded corners=2pt},
  legend cell align=left]
\addplot[jevsubtle, dashed, forget plot] coordinates {(0,0) (34,0)};
\addplot[jevblue, thick, mark=*, mark size=2pt] coordinates {(5.12,-10.17) (8.00,-12.82) (11.96,-1.32) (25.81,-1.74) (31.27,2.56)};
\addlegendentry{Gemma~4, MMEB-V2 image (hit@1)}
\addplot[jevpurple, thick, mark=square*, mark size=2pt] coordinates {(5.12,-10.33) (8.00,-0.60) (11.96,2.61) (25.81,2.94) (31.27,1.65)};
\addlegendentry{Gemma~4, MMEB-V2 visdoc (nDCG@5)}
\addplot[jevrose, thick, mark=triangle*, mark size=2pt] coordinates {(5.12,-10.80) (8.00,-1.12) (11.96,2.61) (25.81,3.39) (31.27,2.41)};
\addlegendentry{Gemma~4, ViDoRe~V3 (nDCG@10)}
\addplot[only marks, mark=diamond, mark size=3pt, thick, jevblue, forget plot] coordinates {(2,-0.36)};
\addplot[only marks, mark=diamond, mark size=3pt, thick, jevpurple, forget plot] coordinates {(2,3.52)};
\addplot[only marks, mark=diamond, mark size=3pt, thick, jevrose, forget plot] coordinates {(2,6.17)};
\addplot[only marks, mark=pentagon, mark size=3pt, thick, jevblue, forget plot] coordinates {(8,3.51)};
\addplot[only marks, mark=pentagon, mark size=3pt, thick, jevpurple, forget plot] coordinates {(8,5.58)};
\addplot[only marks, mark=pentagon, mark size=3pt, thick, jevrose, forget plot] coordinates {(8,8.97)};
\addplot[only marks, mark=star, mark size=3.2pt, thick, jevblue, forget plot] coordinates {(27,1.91)};
\addplot[only marks, mark=star, mark size=3.2pt, thick, jevpurple, forget plot] coordinates {(27,5.59)};
\addplot[only marks, mark=star, mark size=3.2pt, thick, jevrose, forget plot] coordinates {(27,7.93)};
\addplot[only marks, mark=x, mark size=3.2pt, thick, jevblue, forget plot] coordinates {(28.2,1.95)};
\addplot[only marks, mark=x, mark size=3.2pt, thick, jevpurple, forget plot] coordinates {(28.2,6.22)};
\addplot[only marks, mark=x, mark size=3.2pt, thick, jevrose, forget plot] coordinates {(28.2,9.02)};
\addlegendimage{only marks, mark=diamond, mark size=3pt, thick, jevmuted}
\addlegendentry{Qwen3-VL-Reranker-2B (dedicated)}
\addlegendimage{only marks, mark=pentagon, mark size=3pt, thick, jevmuted}
\addlegendentry{Qwen3-VL-Reranker-8B (dedicated)}
\addlegendimage{only marks, mark=star, mark size=3.2pt, thick, jevmuted}
\addlegendentry{Qwen3.8-27B, choice}
\addlegendimage{only marks, mark=x, mark size=3.2pt, thick, jevmuted}
\addlegendentry{JEV-27B-VL, choice}
\end{axis}
\end{tikzpicture}
\caption{Gain over the first stage (top 16 of Qwen3-VL-Embedding-2B), in points of each benchmark's metric, against
model size. Lines: the choice readout on the five Gemma~4 models, at their total parameters. Markers: the rerankers of
the main setup (\cref{tab:main}) at their nominal size, coloured by benchmark; JEV-27B-VL is drawn slightly to the
right of Qwen3.8-27B for readability. Per-task Gemma~4 values: \cref{tab:gemma-image,tab:gemma-visdoc,tab:gemma-vidore}.}
\label{fig:gemma}
\end{figure}
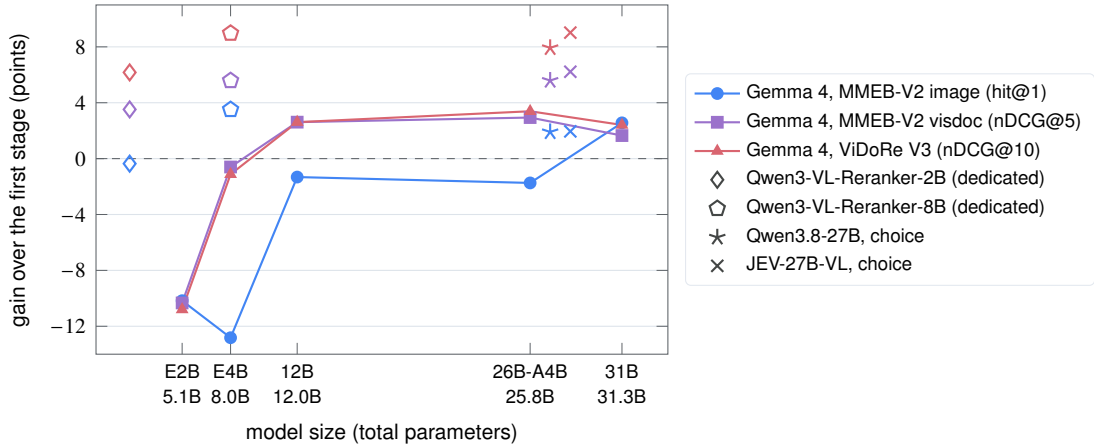

We keep the main setup and replace the backbone with the five instruction-tuned Gemma~4
models~\cite{gemmateam2026gemma4}, E2B, E4B, 12B, 26B-A4B and 31B, applying the choice readout unchanged and with no
training. Gemma's chat template with thinking disabled also ends in an empty thinking block, and the labels A--P are
single tokens in its tokenizer. Two other things change with the backbone: Gemma~4 is a different model family, and
we keep its processor's default of 280 visual tokens per image (the model supports up to
1120~\cite{gemmateam2026gemma4}), against up to 1800 for Qwen3.8-27B (about 4.4k against 28.8k input tokens per
ViDoRe~V3 query). Within the Gemma~4 family the image budget is fixed; against Qwen3.8-27B,
backbone, image budget and size all change. \Cref{fig:gemma} places each Gemma~4 model at its total parameters, 5.1B
to 31.3B; 26B-A4B is a mixture-of-experts model, so its total overstates the computation per token.

\Cref{fig:gemma} shows the results. E2B and E4B lower the first stage on all three benchmarks. This is consistent with ZipRerank's zero-shot
baselines~\cite{sun2026ziprerank}. From 12B upward the readout improves
visual-document retrieval, by 1.7 to 2.9 nDCG@5 on MMEB-V2 visual documents and by 2.4 to 3.4 nDCG@10 on ViDoRe~V3;
on natural images only 31B improves the first stage, by 2.6 hit@1. The gains do not grow monotonically with size: 31B
is the strongest Gemma model on images but is below 26B-A4B on both visual-document benchmarks. Against the rerankers
of the main setup, every Gemma~4 model is below Reranker-2B on both visual-document benchmarks, while Gemma~4 31B is
above Reranker-2B and Qwen3.8-27B on natural images.

\subsection{Efficiency}
\label{sec:efficiency}

\begin{table}[h]
\centering\small
\caption{Latency per query on one H200 at concurrency 1, for the same top-16 ViDoRe~V3 pages (100 English queries from
two domains). Input tokens are the mean over the query's model calls together.}
\label{tab:latency}
\begin{tabular}{lccc}
\toprule
Reranker & model calls & input tokens / query & seconds / query \\
\midrule
Qwen3-VL-Reranker-2B & 16 & 29.6k & 2.75 \\
Qwen3-VL-Reranker-8B & 16 & 29.6k & 3.67 \\
\midrule
Qwen3.8-27B, choice (no decision tuning) & 1 & 28.8k & 3.97 \\
JEV-27B-VL, choice (text decision tuning) & 1 & 28.9k & 4.48 \\
\midrule
Gemma-4-E2B, choice & 1 & 4.4k & 0.47 \\
Gemma-4-E4B, choice & 1 & 4.4k & 0.49 \\
Gemma-4-12B, choice & 1 & 4.4k & 0.55 \\
Gemma-4-26B-A4B, choice & 1 & 4.4k & 0.78 \\
Gemma-4-31B, choice & 1 & 4.4k & 1.27 \\
\bottomrule
\end{tabular}

\end{table}

We keep the main setup and measure latency instead of quality, on 100 English ViDoRe~V3 queries from two domains with
the same top-16 pages for every reranker. Requests run one at a time on one H200, and each query is timed from request
to scores, preprocessing included. \Cref{tab:latency} gives the results. One choice pass of Qwen3.8-27B takes
3.97\,s, 1.1 times the 3.67\,s of Reranker-8B, whose released code scores the 16 pairs one after another, and 1.4
times the 2.75\,s of Reranker-2B. Scoring the 16 pairs in one batch brings Reranker-2B and -8B down to 1.68 and
2.61\,s, and the choice pass then takes 1.5 times as long as Reranker-8B. The three read 28.8k to 29.6k input tokens per query, since each sees the same 16
pages at the same image budget. JEV-27B-VL, with its adapter and decision head, takes 4.48\,s, and one choice pass of
a Gemma~4 model takes 0.47 to 1.27\,s.

\section{Discussion}
\label{sec:discussion}

\subsection{Where the choice helps}
\label{sec:an-query}
Finding~1 rests on benchmark means. The choice readout on Qwen3.8-27B improves the first stage by 1.9 hit@1 on natural
images, by 5.6 nDCG@5 on MMEB-V2 visual documents and by 7.9 nDCG@10 on ViDoRe~V3 (\cref{tab:main}). Over the 12 image tasks (\cref{tab:pertask-image}) it gains on all nine tasks
whose query contains text: most on CIRR (+7.1 hit@1), whose queries combine an image with a modification text, then on
Wiki-SS-NQ (+3.8), VisDial (+3.3) and FashionIQ (+3.1), and by 1.6 to 2.8 points on the other five. It loses on the
three tasks whose query is an image alone: MSCOCO image-to-text ($-4.1$), VisualNews image-to-text ($-1.7$) and NIGHTS
($-0.7$); two of them retrieve captions and one retrieves images. JEV-27B-VL loses on the same three tasks. MM-Embed
found that zero-shot pointwise reranking with a True/False readout helps multimodal queries and lowers accuracy on
single-modal queries~\cite{lin2025mmembed}. With the choice readout, single-modal text queries gain as well, and the
losses are confined to image-only queries.

\subsection{Why the choice beats the yes/no format}
\label{sec:why-choice}
The choice compares the 16 candidates in one context, and their probabilities sum to one. The yes/no readout scores each
candidate in a separate prompt, and nothing places the 16 probabilities on a shared scale. We hypothesize that this is
why the zero-shot yes/no readout falls below the first stage: ranking needs scores that are comparable across
candidates, and supervised pointwise rerankers such as Qwen3-VL-Reranker are trained for reranking with this readout,
while a general VLM is not. The same ordering of the two question types appears for
short videos in a recommendation feed: the developers of JEV-27B-VL report that, on 200 cases built from public
playlists, each with 20 candidates, one choice question per batch reached AUC 0.931 against 0.905 for one yes/no question per candidate~\cite{autotrust2026rankingteam}.

\section{Alternative Views}
\label{sec:alt}

We state the views that most directly challenge the position or the findings that support it, and how far our results
answer them.

\paragraph{AV1: Stronger VLMs make stronger dedicated rerankers too}
Qwen3-VL-Reranker is trained from Qwen3-VL, and a stronger VLM is also a stronger backbone for a dedicated reranker.
Our comparison sets a zero-shot 27B model against supervised 2B and 8B models of a different generation, so it changes
supervision together with scale and generation. It shows that a zero-shot VLM can match a dedicated 8B reranker on MMEB-V2 visual documents today,
not that it would match a dedicated reranker trained from the same backbone; JEV-27B-VL, the best reranker on both
visual-document benchmarks, inherits the same confound. This view limits the practical value of the general route;
\cref{sec:test} describes the test.

\paragraph{AV2: A general VLM is too expensive}
This view qualifies Finding 1. Qwen3.8-27B has more than three times the parameters of Reranker-8B and needs more GPU
memory, and its latency per query is 1.1 times that of Reranker-8B with the released code and 1.5 times with batched
scoring (\cref{sec:efficiency}). In favor of the position, as argued in \cref{sec:intro}, a system that already
serves a VLM could rerank with the same model.

\paragraph{AV3: Small models cannot do it}
This view also qualifies Finding 1. Small general VLMs lower the first stage (\cref{sec:an-scale}), so Finding 1 holds
only for sufficiently large models, and at small sizes a dedicated model is ahead today: Reranker-2B improves both
visual-document benchmarks, where Gemma~4 E2B and E4B lower them.

\paragraph{AV4: Dedicated listwise rerankers already exist}
This view questions what is new in the position. ZipRerank~\cite{sun2026ziprerank} trains for reranking with the
readout we use (\cref{sec:intro}). Our position is that decision tuning can also produce a good reranker, and the two routes
can coexist. Under ours, a model tuned on many kinds of multimodal decisions, of which reranking is one,
should approach a dedicated listwise reranker on reranking while keeping its other decisions.

\paragraph{AV5: The readout has practical limits}
This view concerns the choice readout. One prompt holds at most 16 candidates under the labels A--P, while the
published Qwen3-VL-Reranker numbers rerank the top 100. Deeper lists could be reranked with sliding
windows~\cite{sun2023rankgpt} or with repeated choices over small sets, as in setwise sorting~\cite{zhuang2024setwise},
at one prefill per window or set; the decision endpoint of JEV-27B-VL also extends its labels beyond P with no
retraining~\cite{autotrust2026jevvl}. Listwise readouts depend on option
order~\cite{zhang2026anyjev,huang2026runoff} and prefer some labels whatever the input~\cite{zhang2026anyjev}, and we
present the candidates in first-stage order only. These limits matter for deployment, but they are properties of the
readout rather than arguments for a dedicated model: listwise LLM rerankers in recommendation show position bias as
well~\cite{bito2026positionbias}.

\section{Testing the Position}
\label{sec:test}

\paragraph{Multimodal decision tuning}
Tune a VLM on decisions of many kinds that include visual options: a ranked list with relevance labels gives choice
questions over images and pages, and two candidates of different relevance give a preference question. The position is
refuted if this tuning does not improve the VLM's reranking, in particular on natural images, where text decision
tuning does not help, and on queries that are an image alone, where the choice readout loses (\cref{sec:an-query}).

\paragraph{The gap on one backbone}
On one VLM at several sizes, compare the zero-shot choice, the same model after multimodal decision tuning, and
a dedicated reranker trained from it. A gap to the dedicated reranker that shrinks with size strengthens the general
route; a gap that persists supports AV1 and limits the route's practical value.

\paragraph{Backbone and image budget}
Run Qwen3.8-27B at the image budget of Gemma~4, or Gemma~4 at a larger one. This tells whether the gap between the two
families in \cref{sec:an-scale} comes from the backbone or from the image budget.

\paragraph{The choice readout}
Measure how the choice probabilities change with the order of the candidates (AV5) and how well they are calibrated,
and correct for order, for example by averaging over cyclic rotations~\cite{zhang2026anyjev}. Order effects that
dominate the ranking would weaken the choice format as the interface of a general reranker.

\section{Conclusion}
\label{sec:conclusion}

Multimodal rerankers are usually dedicated models. We argued that a general VLM can serve as a multimodal reranker
through System One decisions, and that decision tuning can improve its reranking. On MMEB-V2 image retrieval, one
choice question over the candidates scores 5.4 to 7.3 hit@1 above a yes/no question per candidate, and the yes/no
question falls below the first stage. With the choice, zero-shot Qwen3.8-27B beats the supervised
Qwen3-VL-Reranker-2B on all three benchmarks, comes within 1.6 points of Reranker-8B on each, and matches it on
MMEB-V2 visual documents, at 1.1 to 1.5 times its latency. The same backbone tuned on text decisions only improves page
reranking by 0.6 to 1.1 points, with gains on most tasks, and leaves natural images unchanged. The zero-shot model is
larger than the supervised rerankers and of a newer generation, so these comparisons are confounded. If the position
holds, multimodal decision tuning offers a route to better multimodal reranking alongside dedicated rerankers;
\cref{sec:test} lists the experiments that would tell.

\bibliographystyle{plainnat}
\bibliography{references,references_manual}

\clearpage
\appendix
\section{Prompts}
\label{app:prompts}

The prompts below are taken from the code of our runs; the released repository sends byte-identical requests. Chat
prompts A and B are shown as Qwen3.8-27B's chat template renders them, with thinking disabled. Each image becomes
``\textless|vision\_start|\textgreater'', its visual tokens and ``\textless|vision\_end|\textgreater''. A query or candidate contributes its images
first, then its text. The task instructions are listed in \cref{tab:instructions}.

\begin{promptbox}{A. Choice prompt, listwise (Qwen3.8-27B)}
<|im_start|>user
Instruction: (*\slot{instruction}*)
Query:(*\slot{query}*)
Candidates:
Candidate A:(*\slot{candidate 1}*)
Candidate B:(*\slot{candidate 2}*)
(*$\;\vdots$*)
Candidate P:(*\slot{candidate 16}*)
Which candidate best matches the query, following the instruction? Answer with the letter of the best candidate only (A-P).<|im_end|>
<|im_start|>assistant
<think>

</think>

(*\readout{next-token logits here, restricted to the label tokens A, \dots, P}*)
\end{promptbox}

\noindent With fewer than 16 candidates, the labels stop at the $K$-th letter and the last line ends with
``(A-$\ell_K$).''

\begin{promptbox}{B. Yes/no prompt, pointwise (Qwen3.8-27B), one per candidate}
<|im_start|>user
Instruction: (*\slot{instruction}*)
Query:(*\slot{query}*)
Candidate:(*\slot{candidate $j$}*)
Does this candidate match the query, following the instruction? Answer only yes or no.<|im_end|>
<|im_start|>assistant
<think>

</think>

(*\readout{next-token logits here, restricted to yes, Yes, no, No}*)
\end{promptbox}

\begin{promptbox}{C. JEV-27B-VL through its decision endpoint (request bodies)}
# choice: one request per query
POST /v1/decide
{"kind": "choice",
 "state": ["Instruction: (*\slot{instruction}*)\nQuery:", (*\slot{query}*), "\nCandidates:",
           "\nCandidate A:", (*\slot{candidate 1}*), ..., "\nCandidate P:", (*\slot{candidate 16}*)],
 "question": "Which candidate best matches the query, following the instruction?",
 "options": ["Candidate A", ..., "Candidate P"],
 "strategy": "single", "thinking": "off"}
(*\readout{response ``probabilities'': one per option, from the decision head}*)

# yes/no: one request per candidate
POST /v1/decide
{"kind": "noul",
 "state": ["Instruction: (*\slot{instruction}*)\nQuery:", (*\slot{query}*), "\nCandidate:", (*\slot{candidate $j$}*)],
 "question": "Does this candidate match the query, following the instruction?",
 "thinking": "off"}
(*\readout{response ``probabilities'' for [false, true]; the second is $P(\mathrm{yes})$}*)
\end{promptbox}

\noindent In the ``state'' field, text parts are strings and each image is an object ``\{"image": [data URL]\}''.

\begin{promptbox}{D. Qwen3-VL-Reranker, released code (messages before the chat template)}
[{"role": "system",
  "content": "Judge whether the Document meets the requirements based on the Query and the Instruct provided. Note that the answer can only be \"yes\" or \"no\"."},
 {"role": "user",
  "content": ["<Instruct>: (*\slot{instruction}*)", "<Query>:", (*\slot{query}*), "\n<Document>:", (*\slot{candidate $j$}*)]}]
(*\readout{chat template with the generation prompt; score $\sigma(z_{\mathrm{yes}} - z_{\mathrm{no}})$ at the last position}*)
\end{promptbox}

\begin{table}[h]
\centering\small
\caption{Task instructions. MMEB-V2 image tasks use the instruction of the benchmark's annotations (one per task),
as loaded by the Qwen3-VL-Embedding evaluation code; visual-document tasks and ViDoRe~V3 share one instruction.}
\label{tab:instructions}
\begin{tabularx}{\linewidth}{lX}
\toprule
Task & Instruction \\
\midrule
CIRR & Given an image, find a similar everyday image with the described changes: \\
EDIS & Find a news image that matches the provided caption: \\
FashionIQ & Find an image to match the fashion image and style note: \\
MSCOCO i2t & Find an image caption describing the given everyday image. \\
MSCOCO t2i & Find me an everyday image that matches the given caption: \\
NIGHTS & Find a day-to-day image that looks similar to the provided image. \\
OVEN & Retrieve a Wikipedia image-description pair that provides evidence for the question of this image: \\
VisDial & Represent the given dialogue about an image, which is used for image retrieval: \\
VisualNews i2t & Find a caption for the news in the given photo. \\
VisualNews t2i & Retrieve an image of this news caption. \\
WebQA & Find a Wikipedia image that answers this question: \\
Wiki-SS-NQ & Find the document image that can answer the given query: \\
\midrule
MMEB-V2 visdoc (24 tasks), ViDoRe~V3 & Find a document image that matches the given query. \\
\bottomrule
\end{tabularx}
\end{table}

\section{Implementation details}
\label{app:impl}

\paragraph{Reading the label probabilities}
Let $x$ be the token sequence of prompt~A, rendered by the chat template with thinking disabled and ending where the
answer starts, and let $z = f_\theta(x) \in \mathbb{R}^{|\mathcal{V}|}$ be the next-token logits at that position. With
label tokens $\ell_1, \dots, \ell_K$ (the single tokens ``A'', ``B'', \dots; ids 32 to 47 in the Qwen3.8
tokenizer), candidate $c_j$ receives $p_j$ by \cref{eq:choice}.
The yes/no readout of prompt~B uses the tokens ``yes'', ``Yes'', ``no'', ``No'' (ids 9405, 9175,
2083, 2665) and scores a candidate by
\begin{equation}
  P(\mathrm{yes}) = \frac{\pi_{\mathrm{yes}} + \pi_{\mathrm{Yes}}}{\pi_{\mathrm{yes}} + \pi_{\mathrm{Yes}} + \pi_{\mathrm{no}} + \pi_{\mathrm{No}}},
  \qquad \pi_w = \exp z_w .
\end{equation}
Both read a single forward pass; nothing is decoded.

\paragraph{Obtaining the logits from vLLM}
Each prompt is one chat-completion request with the fields of \cref{tab:request}. The server runs with
``-{}-logprobs-mode processed\_logprobs'', so the returned log-probabilities are those of the distribution after
the allowed-token mask, and the explicit sampling parameters keep the model's default sampling settings from
truncating it. The client applies a softmax to the $K$ returned values, which gives \cref{eq:choice}; the sampled token
is not used.

\begin{table}[h]
\centering\small
\caption{The chat-completion request behind prompts A and B.}
\label{tab:request}
\begin{tabularx}{\linewidth}{lX}
\toprule
Field & Value \\
\midrule
``model'' & ``autotrust/JEV-27B-VL'' (served with the unmodified Qwen3.8-27B weights) \\
``messages'' & one user turn with the parts of prompt A or B \\
``max\_tokens'' & 1 \\
``allowed\_token\_ids'' & the label tokens: A, \dots, $\ell_K$ (ids 32--47), or yes, Yes, no, No \\
``logprobs'', ``top\_logprobs'' & true, the number of label tokens \\
``return\_tokens\_as\_token\_ids'' & true \\
``temperature'', ``top\_p'', ``top\_k'' & 1, 1, 0 \\
``chat\_template\_kwargs'' & ``\{"enable\_thinking": false\}'' \\
\bottomrule
\end{tabularx}
\end{table} In all 14{,}419 choice requests to Qwen3.8-27B, every one of the 230{,}704 labels received a
log-probability.

\paragraph{JEV-27B-VL}
Its developers train the LoRA adapter and the decision head on the ``jev-distill-corpus-v3'' dataset of text
decisions; the tuned head answers ``noul'' questions, ``choice'' questions with 2 to 256 options and 0--5 ratings. On
held-out rows of that dataset labelled by TypeSafe's Jev~1.13, the model card reports a KL divergence of about 0.017
from Jev's output distributions. The endpoint ``/v1/decide'' takes the state, the question and, for ``choice'', the
options, and returns one probability per option from one forward pass. The same state and question as for
Qwen3.8-27B go to this endpoint (prompt~C). Its decision head returns one probability
per option, and these probabilities are $p_j$. For the yes/no readout, a request whose question type (the field ``kind'') is ``noul'' returns
probabilities for ``false'' and ``true''; we use the second.

\paragraph{Ranking}
Candidates are sorted by $p_j$ in decreasing order, ties broken by first-stage rank, and candidates below the top $K$
keep their first-stage rank. A query whose request fails would keep the first-stage order; no request failed in any run
reported here.

\paragraph{Inputs}
Images are converted to RGB and, when larger than $1800 \times 32 \times 32 = 1{,}843{,}200$ pixels, downscaled with
bicubic resampling at the same aspect ratio; they are sent as JPEG data URLs (quality 90), and the server's image
processor then resizes them as it does any input. Runs of whitespace in text are collapsed to one space. The choice
prompts are about 3.1k tokens long on MMEB-V2 image retrieval, 17.8k on visual-document retrieval and 28.6k on
ViDoRe~V3.

\paragraph{Serving}
Both 27B models are served by the server in the JEV-27B-VL repository (model revision fee8c1d) on
vLLM~0.30.0, started as below. Chat requests for the served name ``autotrust/\allowbreak JEV-27B-VL'' use the unmodified
Qwen3.8-27B weights; ``/v1/decide'' adds the LoRA adapter and the decision head. The client keeps 8 queries in
flight; the yes/no readout sends the 16 requests of a query one after another.

\begin{promptbox}{Server command (one H200)}
python3 JEV-27B-VL/serve_decide.py --model JEV-27B-VL --served-model-name autotrust/JEV-27B-VL \
  --enable-lora --max-lora-rank 32 --lora-modules jev-decision=JEV-27B-VL/adapter_vllm \
  --logprobs-mode processed_logprobs --max-model-len 65536 --enable-prefix-caching \
  --mamba-cache-mode align --max-num-seqs 8 --trust-request-chat-template \
  --gpu-memory-utilization 0.92 --limit-mm-per-prompt '{"image": 80}'   # 20 on ViDoRe V3
\end{promptbox}

\paragraph{Supervised rerankers}
Qwen3-VL-Reranker-2B (revision 4bd860a) and -8B (b212dc8) run with the released code of the
Qwen3-VL-Embedding repository at commit 393e297, in bf16 with FlashAttention-2: on MMEB-V2 through the
repository's reranker evaluation, and on ViDoRe~V3 through the same reranker class (prompt~D). The class scores the 16
pairs of a query one after another, truncates each pair to 10{,}240 tokens and caps each image at $1800 \times 32
\times 32$ pixels. Its score is $\sigma\big((w_{\mathrm{yes}} - w_{\mathrm{no}})^{\top} h\big)$, with $h$ the last hidden
state and $w$ rows of the language-model head, which equals $\sigma(z_{\mathrm{yes}} - z_{\mathrm{no}})$. Under
transformers~5.8, a one-line patch rebuilds the processor's ``mm\_token\_type\_ids'' from the final input ids; under
transformers~4.57 it changes nothing.

\paragraph{First stage}
Qwen3-VL-Embedding-2B (revision 9f2f7e7). On MMEB-V2 we run the repository's embedding evaluation, whose
rankings its reranker evaluation reads. On ViDoRe~V3 we use the same embedding class: pages with its default
instruction, queries with the instruction of \cref{tab:instructions}, and the top 100 pages by the inner product of the
returned embeddings.

\paragraph{Metrics}
MMEB-V2 uses the repository's ranking metrics: hit@1 for image retrieval and ``ndcg\_linear@5'' for visual
documents, averaged over tasks. For ViDoRe~V3, with graded relevance $g$ from the qrels and ranking $r$,
$\mathrm{nDCG@10} = \mathrm{DCG}/\mathrm{IDCG}$, where $\mathrm{DCG} = \sum_{i=1}^{10} g(r_i)/\log_2(i+1)$ and IDCG is
the same sum over the ten largest positive gains. It is computed per dataset over the queries of all six languages and
averaged over the eight datasets.

\paragraph{Hardware and software}
One NVIDIA H200 per run. The Qwen models run with torch~2.8, transformers~5.8 and flash-attn~2.7.3; the 27B models run
on vLLM~0.30.0.

\section{Per-task results}
\label{app:pertask}

\begin{table}[h]
\centering\small\setlength{\tabcolsep}{4.5pt}
\caption{MMEB-V2 image retrieval per task (hit@1, top-16 of Qwen3-VL-Embedding-2B). Bold: best reranker in the row, underlined: second best. JEV-27B-VL (text decision tuning) and Qwen3.8-27B (no decision tuning): choice readout; the two yes/no columns (y/n) are the readout comparison of \cref{tab:2x2}.}
\label{tab:pertask-image}
\begin{tabular}{lccccccc}
\toprule
Task & Emb-2B & R-2B & R-8B & JEV-27B-VL & Qwen3.8-27B & Qwen3.8 y/n & JEV y/n \\
\midrule
CIRR & 64.90 & 65.60 & \textbf{74.10} & 71.80 & \underline{72.00} & 66.00 & 68.00 \\
EDIS & 92.50 & 87.40 & 92.40 & \underline{94.10} & \textbf{95.00} & 83.80 & 87.10 \\
FashionIQ & 38.80 & 41.10 & 41.20 & \underline{41.40} & \textbf{41.90} & 32.50 & 38.10 \\
MSCOCO i2t & 74.60 & \underline{73.90} & \textbf{77.70} & 73.80 & 70.50 & 62.10 & 62.80 \\
MSCOCO t2i & 77.30 & 79.80 & \textbf{82.20} & \underline{80.20} & 80.00 & 73.10 & 75.70 \\
NIGHTS & 69.50 & 67.70 & \textbf{70.80} & 68.10 & \underline{68.80} & 64.10 & 67.10 \\
OVEN & 76.40 & 76.10 & \textbf{78.60} & 76.50 & \underline{78.00} & 70.30 & 70.20 \\
VisDial & 81.50 & 84.80 & \textbf{89.50} & \underline{87.20} & 84.80 & 80.10 & 84.80 \\
VisualNews i2t & 75.90 & 72.70 & \textbf{79.60} & 74.00 & \underline{74.20} & 62.70 & 62.00 \\
VisualNews t2i & 73.40 & 68.90 & 75.10 & \underline{75.60} & \textbf{76.20} & 67.30 & 70.40 \\
WebQA & 89.90 & 91.30 & \textbf{92.70} & 91.10 & \underline{92.40} & 89.00 & 87.40 \\
Wiki-SS-NQ & 83.50 & 84.60 & 86.40 & \textbf{87.80} & \underline{87.30} & 83.00 & 83.50 \\
\midrule
Mean & 74.85 & 74.49 & \textbf{78.36} & \underline{76.80} & 76.76 & 69.50 & 71.42 \\
\bottomrule
\end{tabular}
\end{table}

\begin{table}[h]
\centering\small\setlength{\tabcolsep}{4.5pt}
\caption{MMEB-V2 visual-document retrieval per task (nDCG@5, top-16 of Qwen3-VL-Embedding-2B). Bold: best reranker in the row, underlined: second best. JEV-27B-VL (text decision tuning) and Qwen3.8-27B (no decision tuning): choice readout.}
\label{tab:pertask-visdoc}
\begin{tabular}{lccccc}
\toprule
Task & Emb-2B & R-2B & R-8B & JEV-27B-VL & Qwen3.8-27B \\
\midrule
MMLongBench-doc & 54.99 & \underline{57.81} & \textbf{58.23} & 57.47 & 55.90 \\
MMLongBench-page & 55.73 & 61.16 & 63.77 & \textbf{66.21} & \underline{65.38} \\
ViDoRe arxivqa & 81.80 & 86.15 & \textbf{89.52} & \underline{87.30} & 85.67 \\
ViDoRe biomedical lectures v2 (multi.) & 68.23 & 66.52 & 69.35 & \textbf{73.41} & \underline{72.78} \\
ViDoRe docvqa & 49.36 & 57.32 & \textbf{58.75} & \underline{58.47} & 57.66 \\
ViDoRe economics reports v2 (multi.) & 63.09 & 60.75 & 67.58 & \textbf{73.58} & \underline{70.81} \\
ViDoRe esg reports (human) v2 & 64.11 & 69.96 & 75.44 & \textbf{83.26} & \underline{81.52} \\
ViDoRe esg reports v2 (multi.) & 65.49 & 63.53 & 71.86 & \textbf{74.57} & \underline{72.09} \\
ViDoRe infovqa & 88.35 & 93.08 & \underline{93.54} & \textbf{93.81} & 93.08 \\
ViDoRe shiftproject & 83.90 & \underline{90.94} & \textbf{91.57} & 90.60 & 89.65 \\
ViDoRe synthDocQA artificial intelligence & 97.05 & \underline{99.63} & \underline{99.63} & \textbf{100.00} & \textbf{100.00} \\
ViDoRe synthDocQA energy & 90.84 & \underline{96.56} & \textbf{96.63} & 96.39 & 96.02 \\
ViDoRe synthDocQA government reports & 95.68 & 96.80 & \textbf{98.52} & \underline{98.15} & \textbf{98.52} \\
ViDoRe synthDocQA healthcare industry & 97.09 & \underline{99.26} & \textbf{100.00} & 97.95 & 99.02 \\
ViDoRe tabfquad & 96.40 & 97.27 & \textbf{98.27} & 97.54 & \underline{97.58} \\
ViDoRe tatdqa & 63.19 & \underline{78.03} & \textbf{80.31} & 77.00 & 76.82 \\
ViDoSeek-doc & 85.00 & \underline{85.75} & \textbf{86.14} & 85.07 & 83.46 \\
ViDoSeek-page & 81.94 & 86.67 & 92.75 & \textbf{95.90} & \underline{95.47} \\
VisRAG ArxivQA & 83.58 & 87.08 & \textbf{89.65} & 87.70 & \underline{88.58} \\
VisRAG ChartQA & 87.13 & \underline{91.89} & 89.55 & \textbf{93.06} & \textbf{93.06} \\
VisRAG InfoVQA & 92.88 & \underline{97.23} & \textbf{97.33} & 96.87 & 96.84 \\
VisRAG MP-DocVQA & 87.06 & 92.51 & 93.35 & \textbf{94.17} & \underline{93.42} \\
VisRAG PlotQA & 72.33 & 71.85 & \textbf{74.97} & 73.47 & \underline{73.74} \\
VisRAG SlideVQA & 95.59 & 97.52 & \underline{98.00} & \textbf{98.08} & 97.83 \\
\midrule
Mean & 79.20 & 82.72 & 84.78 & \textbf{85.42} & \underline{84.79} \\
\bottomrule
\end{tabular}
\end{table}

\begin{table}[h]
\centering\small\setlength{\tabcolsep}{4.5pt}
\caption{ViDoRe~V3 per dataset (nDCG@10, all six languages pooled; number of queries in parentheses). Bold: best reranker in the row, underlined: second best. JEV-27B-VL (text decision tuning) and Qwen3.8-27B (no decision tuning): choice readout.}
\label{tab:pertask-vidore}
\begin{tabular}{lccccc}
\toprule
Dataset & Emb-2B & R-2B & R-8B & JEV-27B-VL & Qwen3.8-27B \\
\midrule
computer science (1,290) & 72.21 & 77.83 & \underline{79.36} & \textbf{80.03} & 78.35 \\
energy (1,848) & 58.06 & 64.90 & 68.29 & \textbf{70.00} & \underline{69.02} \\
finance en (1,854) & 56.47 & 63.25 & \textbf{66.22} & \underline{66.08} & 64.38 \\
finance fr (1,920) & 36.25 & 44.44 & \textbf{47.48} & \underline{47.41} & 46.03 \\
hr (1,908) & 51.96 & 59.34 & \underline{63.21} & \textbf{63.64} & 62.21 \\
industrial (1,698) & 40.32 & 48.44 & \textbf{51.38} & \underline{50.73} & 50.08 \\
pharmaceuticals (2,184) & 61.94 & 65.42 & \underline{68.20} & \textbf{68.38} & 67.66 \\
physics (1,812) & 46.86 & 49.83 & \textbf{51.69} & \underline{49.97} & 49.81 \\
\midrule
Mean & 53.01 & 59.18 & \underline{61.98} & \textbf{62.03} & 60.94 \\
\bottomrule
\end{tabular}
\end{table}

The Gemma~4 results of \cref{fig:gemma}, per task and per dataset:

\begin{table}[h]
\centering\small\setlength{\tabcolsep}{3.5pt}
\caption{Gemma~4 on MMEB-V2 image retrieval per task (hit@1, top-16 of Qwen3-VL-Embedding-2B, choice readout). Emb-2B is the first stage and Qwen3.8 the choice readout on Qwen3.8-27B; bold and underline mark the best and second best of the five Gemma sizes.}
\label{tab:gemma-image}
\begin{tabular}{lccccccc}
\toprule
Task & Emb-2B & Qwen3.8 & E2B & E4B & 12B & 26B-A4B & 31B \\
\midrule
CIRR & 64.90 & 72.00 & 60.00 & 35.50 & \underline{71.30} & 68.10 & \textbf{75.80} \\
EDIS & 92.50 & 95.00 & 91.40 & 78.00 & 85.70 & \underline{93.90} & \textbf{96.00} \\
FashionIQ & 38.80 & 41.90 & 22.10 & 19.70 & \underline{41.20} & 35.90 & \textbf{43.10} \\
MSCOCO i2t & 74.60 & 70.50 & 46.20 & 63.70 & \underline{72.50} & 69.00 & \textbf{74.30} \\
MSCOCO t2i & 77.30 & 80.00 & 75.50 & 68.20 & \underline{79.00} & 78.20 & \textbf{80.10} \\
NIGHTS & 69.50 & 68.80 & \textbf{68.10} & 57.80 & \underline{67.40} & 59.70 & 65.80 \\
OVEN & 76.40 & 78.00 & \underline{73.70} & 59.80 & 67.40 & 67.30 & \textbf{74.70} \\
VisDial & 81.50 & 84.80 & 75.40 & 73.80 & \underline{82.00} & 81.20 & \textbf{85.60} \\
VisualNews i2t & 75.90 & 74.20 & 34.30 & 57.60 & 69.70 & \underline{71.90} & \textbf{77.70} \\
VisualNews t2i & 73.40 & 76.20 & 71.70 & 66.50 & 74.20 & \underline{77.30} & \textbf{79.50} \\
WebQA & 89.90 & 92.40 & 78.40 & 84.20 & 88.60 & \textbf{91.50} & \underline{91.40} \\
Wiki-SS-NQ & 83.50 & 87.30 & 79.30 & 79.60 & \underline{83.40} & 83.30 & \textbf{84.90} \\
\midrule
Mean & 74.85 & 76.76 & 64.68 & 62.03 & \underline{73.53} & 73.11 & \textbf{77.41} \\
\bottomrule
\end{tabular}
\end{table}

\begin{table}[h]
\centering\small\setlength{\tabcolsep}{3.5pt}
\caption{Gemma~4 on MMEB-V2 visual-document retrieval per task (nDCG@5, top-16 of Qwen3-VL-Embedding-2B, choice readout). Emb-2B is the first stage and Qwen3.8 the choice readout on Qwen3.8-27B; bold and underline mark the best and second best of the five Gemma sizes.}
\label{tab:gemma-visdoc}
\begin{tabular}{lccccccc}
\toprule
Task & Emb-2B & Qwen3.8 & E2B & E4B & 12B & 26B-A4B & 31B \\
\midrule
MMLongBench-doc & 54.99 & 55.90 & 48.84 & \underline{54.67} & 54.27 & \textbf{54.78} & 49.55 \\
MMLongBench-page & 55.73 & 65.38 & 44.40 & 57.41 & \underline{61.47} & \textbf{61.87} & 60.26 \\
ViDoRe arxivqa & 81.80 & 85.67 & 73.86 & 81.25 & \textbf{84.84} & 84.08 & \underline{84.61} \\
ViDoRe biomedical lectures v2 (multi.) & 68.23 & 72.78 & 57.62 & 65.05 & \textbf{70.32} & \underline{67.77} & 63.90 \\
ViDoRe docvqa & 49.36 & 57.66 & 45.05 & 51.67 & 55.00 & \textbf{56.42} & \underline{55.38} \\
ViDoRe economics reports v2 (multi.) & 63.09 & 70.81 & 52.44 & 61.17 & \textbf{65.69} & \underline{64.53} & 61.18 \\
ViDoRe esg reports (human) v2 & 64.11 & 81.52 & 57.18 & 71.83 & 72.13 & \textbf{76.90} & \underline{74.25} \\
ViDoRe esg reports v2 (multi.) & 65.49 & 72.09 & 56.00 & 66.88 & \underline{68.98} & \textbf{71.80} & 65.21 \\
ViDoRe infovqa & 88.35 & 93.08 & 81.81 & 87.78 & 90.33 & \textbf{92.02} & \underline{91.38} \\
ViDoRe shiftproject & 83.90 & 89.65 & 48.74 & 75.80 & \underline{84.56} & 84.47 & \textbf{85.66} \\
ViDoRe synthDocQA artificial intelligence & 97.05 & 100.00 & 86.41 & 94.31 & \underline{98.26} & 97.79 & \textbf{99.63} \\
ViDoRe synthDocQA energy & 90.84 & 96.02 & 84.38 & 91.15 & \underline{94.07} & \textbf{94.10} & 93.35 \\
ViDoRe synthDocQA government reports & 95.68 & 98.52 & 84.79 & 92.44 & \underline{96.82} & 96.48 & \textbf{97.02} \\
ViDoRe synthDocQA healthcare industry & 97.09 & 99.02 & 86.37 & 92.32 & \textbf{96.91} & 94.07 & \underline{96.43} \\
ViDoRe tabfquad & 96.40 & 97.58 & 75.22 & 96.30 & \textbf{97.64} & \underline{97.49} & 97.39 \\
ViDoRe tatdqa & 63.19 & 76.82 & 46.29 & 61.28 & 66.40 & \textbf{69.34} & \underline{68.15} \\
ViDoSeek-doc & 85.00 & 83.46 & 79.76 & 81.54 & \textbf{82.25} & \underline{82.00} & 73.33 \\
ViDoSeek-page & 81.94 & 95.47 & 81.21 & 90.71 & \textbf{94.62} & 93.82 & \underline{94.15} \\
VisRAG ArxivQA & 83.58 & 88.58 & 76.09 & 83.21 & \underline{86.59} & 85.73 & \textbf{87.52} \\
VisRAG ChartQA & 87.13 & 93.06 & 81.13 & 88.23 & \underline{91.48} & \textbf{91.69} & 91.37 \\
VisRAG InfoVQA & 92.88 & 96.84 & 80.42 & 92.33 & 93.71 & \underline{95.10} & \textbf{95.79} \\
VisRAG MP-DocVQA & 87.06 & 93.42 & 74.98 & 87.71 & 90.27 & \underline{91.31} & \textbf{91.90} \\
VisRAG PlotQA & 72.33 & 73.74 & 63.39 & 66.56 & \underline{70.69} & \textbf{71.03} & 69.46 \\
VisRAG SlideVQA & 95.59 & 97.83 & 86.55 & 94.82 & \underline{96.09} & \textbf{96.68} & 93.60 \\
\midrule
Mean & 79.20 & 84.79 & 68.87 & 78.60 & \underline{81.81} & \textbf{82.14} & 80.85 \\
\bottomrule
\end{tabular}
\end{table}

\begin{table}[h]
\centering\small\setlength{\tabcolsep}{3.5pt}
\caption{Gemma~4 on ViDoRe~V3 per dataset (nDCG@10, all six languages pooled, choice readout). Columns and marks as in \cref{tab:gemma-image}.}
\label{tab:gemma-vidore}
\begin{tabular}{lccccccc}
\toprule
Dataset & Emb-2B & Qwen3.8 & E2B & E4B & 12B & 26B-A4B & 31B \\
\midrule
computer science & 72.21 & 78.35 & 55.37 & 69.33 & \textbf{74.48} & \underline{74.34} & 70.66 \\
energy & 58.06 & 69.02 & 40.64 & 56.59 & 61.66 & \textbf{63.96} & \underline{63.73} \\
finance en & 56.47 & 64.38 & 45.67 & 53.83 & 58.46 & \textbf{59.10} & \underline{58.50} \\
finance fr & 36.25 & 46.03 & 26.39 & 36.11 & 39.64 & \textbf{41.22} & \underline{41.14} \\
hr & 51.96 & 62.21 & 41.90 & 49.87 & 54.11 & \textbf{55.74} & \underline{55.57} \\
industrial & 40.32 & 50.08 & 33.08 & 40.81 & 44.40 & \underline{45.63} & \textbf{45.77} \\
pharmaceuticals & 61.94 & 67.66 & 55.48 & 61.80 & \textbf{64.61} & \underline{64.43} & 62.49 \\
physics & 46.86 & 49.81 & 39.14 & \underline{46.78} & \textbf{47.58} & 46.76 & 45.51 \\
\midrule
Mean & 53.01 & 60.94 & 42.21 & 51.89 & \underline{55.62} & \textbf{56.40} & 55.42 \\
\bottomrule
\end{tabular}
\end{table}

\end{document}